\PassOptionsToPackage{unicode}{hyperref}
\PassOptionsToPackage{hyphens}{url}
\documentclass[
]{article}
\usepackage{amsmath,amssymb}
\usepackage{lmodern}
\usepackage{ifxetex,ifluatex}
\ifnum 0\ifxetex 1\fi\ifluatex 1\fi=0 
  \usepackage[T1]{fontenc}
  \usepackage[utf8]{inputenc}
  \usepackage{textcomp} 
\else 
  \usepackage{unicode-math}
  \defaultfontfeatures{Scale=MatchLowercase}
  \defaultfontfeatures[\rmfamily]{Ligatures=TeX,Scale=1}
\fi
\IfFileExists{upquote.sty}{\usepackage{upquote}}{}
\IfFileExists{microtype.sty}{
  \usepackage[]{microtype}
  \UseMicrotypeSet[protrusion]{basicmath} 
}{}
\makeatletter
\@ifundefined{KOMAClassName}{
  \IfFileExists{parskip.sty}{%
    \usepackage{parskip}
  }{
    \setlength{\parindent}{0pt}
    \setlength{\parskip}{6pt plus 2pt minus 1pt}}
}{
  \KOMAoptions{parskip=half}}
\makeatother
\usepackage{xcolor}
\IfFileExists{xurl.sty}{\usepackage{xurl}}{} 
\IfFileExists{bookmark.sty}{\usepackage{bookmark}}{\usepackage{hyperref}}
\hypersetup{
  pdftitle={Mapping scientific communities at scale},
  pdfkeywords={scanR, VOSviewer, graphology, scientific
ccommunity, community detection, research portal, Elasticsearch, network
analysis, network vizualization},
  hidelinks,
  pdfcreator={LaTeX via pandoc}}
\usepackage[left=3cm, right=3cm, top=3cm, bottom=3cm]{geometry}
\usepackage{color}
\usepackage{fancyvrb}

\DefineVerbatimEnvironment{Highlighting}{Verbatim}{commandchars=\\\{\}}
\newenvironment{Shaded}{}{}

\newcommand{\DataTypeTok}[1]{\textcolor[rgb]{0.56,0.13,0.00}{#1}}
\newcommand{\DecValTok}[1]{\textcolor[rgb]{0.25,0.63,0.44}{#1}}

\newcommand{\ErrorTok}[1]{\textcolor[rgb]{1.00,0.00,0.00}{\textbf{#1}}}

\newcommand{\FunctionTok}[1]{\textcolor[rgb]{0.02,0.16,0.49}{#1}}

\newcommand{\StringTok}[1]{\textcolor[rgb]{0.25,0.44,0.63}{#1}}

\usepackage{graphicx}
\makeatletter
\def\maxwidth{\ifdim\Gin@nat@width>\linewidth\linewidth\else\Gin@nat@width\fi}
\def\maxheight{\ifdim\Gin@nat@height>\textheight\textheight\else\Gin@nat@height\fi}
\makeatother
\setkeys{Gin}{width=\maxwidth,height=\maxheight,keepaspectratio}
\makeatletter
\def\fps@figure{htbp}
\makeatother
\ifluatex
  \usepackage{selnolig}  
\fi
\newlength{\cslhangindent}
\newlength{\csllabelwidth}
 {
  \setlength{\parindent}{0pt}
  \ifodd #1 \everypar{\setlength{\hangindent}{\cslhangindent}}\ignorespaces\fi
  \ifnum #2 > 0
  \setlength{\parskip}{#2\baselineskip}
  \fi
 }%
 {}
\usepackage{calc}

\newenvironment{cslreferences}%
  {\setlength{\parindent}{0pt}%
  \everypar{\setlength{\hangindent}{\cslhangindent}}\ignorespaces}%
  {\par}

\title{Mapping scientific communities at scale}
\usepackage{authblk}
\author[%
  2%
  ]{%
  Victor Barbier%
}
\author[%
  1%
  ]{%
  Eric Jeangirard%
}
\affil[1]{French Ministry of Higher Education and Research, Paris,
France}
\affil[2]{National Institute for Research in Digital Science and
Technology, INRIA, Paris, France}
\date{January 2025}

\makeatletter
\def\@maketitle{%
  \newpage \null \vskip 2em
  \begin {center}%
    \let \footnote \thanks
         {\LARGE \@title \par}%
         \vskip 1.5em%
                {\large \lineskip .5em%
                  \begin {tabular}[t]{c}%
                    \@author
                  \end {tabular}\par}%
                                                \vskip 1em{\large \@date}%
  \end {center}%
  \par
  \vskip 1.5em}
\makeatother

\begin{document}
\maketitle
\begin{abstract}
This study introduces a novel methodology for mapping scientific
communities at scale, addressing challenges associated with network
analysis in large bibliometric datasets. By leveraging enriched
publication metadata from the French research portal scanR and applying
advanced filtering techniques to prioritize the strongest interactions
between entities, we construct detailed, scalable network maps. These
maps are enhanced through systematic disambiguation of authors,
affiliations, and topics using persistent identifiers and specialized
algorithms. The proposed framework integrates Elasticsearch for
efficient data aggregation, Graphology for network spatialization (Force
Atltas2) and community detection (Louvain algorithm) and VOSviewer for
network vizualization. A Large Language Model (Mistral Nemo) is used to
label the communities detected and OpenAlex data helps to enrich the
results with citation counts estimation to detect hot topics. This
scalable approach enables insightful exploration of research
collaborations and thematic structures, with potential applications for
strategic decision-making in science policy and funding. These web tools
are effective at the global (national) scale but are also available (and
can be integrated via iframes) on the perimeter of any French research
institution (from large research organisms to any laboratory). The scanR
community analysis tool is available online
\url{https://scanr.enseignementsup-recherche.gouv.fr/networks/get-started}.
All tools and methodologies are open-source on the repo
\url{https://github.com/dataesr/scanr-ui}.
\end{abstract}

\textbf{Keywords}: scanR, VOSviewer, graphology, scientific community,
community detection, research portal, Elasticsearch, network analysis,
network vizualization

\hypertarget{motivation}{%
\section{1. Motivation}\label{motivation}}

Analysing and mapping scientific communities provides an insight into
the structure and evolution of academic disciplines. This involves
providing an analytical and visual representation of the relationships
between entities (e.g.~researchers, research laboratories, research
themes), with the aim, in particular, of understanding the networks and
dynamics of scientific collaboration, and identifying collaborative
groups and their influences. From the point of view of decision-makers,
this type of tool is useful for strategic decision-making with a view to
public policy and funding.

These maps are generally deduced from data in bibliographic databases
(open or proprietary), based on co-publication or citation information.
In the case of co-publications, two entities (authors, for example) will
be linked if they have collaborated (co-published) on a piece of
research. These links are then symmetrical (bi-directional). In the case
of citation links, two authors will be linked if one cites the research
work of another, in the list of references. This is a directed link, as
one author may cite another without this being reciprocal. Many recent
works use this second approach, for example by trying to calculate
composite indicators of novelty (or innovation) based on citation links
(Wang and Barabási 2021).

The quality and completeness of the bibliographic metadata used are
essential if we are to produce a relevant map. Today, the quality of
open citation data still needs to be improved (Alperin et al. 2024). On
the other hand, it is possible to obtain quality metadata on
publications (and therefore links to co-publications). For example, the
French Open Science Monitor (BSO) has compiled a corpus of French
publications with good coverage (Chaignon and Egret 2022). This corpus
is exposed in the French research portal scanR (Jeangirard 2024). This
is a corpus containing about 4 millions publications in all disciplines.
These publications have been enriched with disambiguated persistent
identifier (PID) on authors, affiliations and topics.

\hypertarget{previous-limits-of-the-scanr-application}{%
\subsection{1.1 Previous limits of the scanR
application}\label{previous-limits-of-the-scanr-application}}

Launched in 2016, the scanR portal used to be a search engine. Its scope
first focused on research entities (institutions, laboratories and
private companies) and was extended in 2020 to cover fundings,
publications, patents and authors. Two main use cases were covered.
Firstly, the ability to generate a list of search results corresponding
to a user query. A list of laboratories, authors, funding or
publications could be generated. Secondly, for each institution (or
laboratory), a unified view of all is data was grouped together on a
dedicated page in scanR (administrative information, list of
publications, list of fundings, main partners, etc.).

However, these functions only gave a flat view of the different
dimensions, without providing any insights into the interactions between
institutions, laboratories or authors. For a user interested in a
research theme, for example, the list of the main contributors (those
who have co-authored the most publications) does not give a clear idea
of which research communities are at work and how they interact with
each other. A network analysis tool to describe these interactions and
attempt to detect research communities could therefore enable us to go
further in creating tools to help explore fields of research and
innovation.

\hypertarget{network-analysis-limits}{%
\subsection{1.2 Network analysis limits}\label{network-analysis-limits}}

Network analysis tools for bibliographic studies are used to study the
relationships between entities in a corpus. In general, the size of this
corpus is limited because the calculations to determine the nodes, links
and their positions for very large networks require too many resources,
in addition to being very difficult to interpret. As a result, tools
such as VOSviewer offer options for limiting the size of networks. The
first option is to filter publications with too many authors. This is
particularly true of publications in particle physics, which can list
several thousand authors. As well as generating very large networks,
this hyperauthorship can also be seen as reducing the relevance of the
information conveyed by the co-authorship links. The second option
offered by VOSviewer is to set thresholds to limit the number of nodes
directly (minimum number of publications or minimum number of citations
for a node). However, this approach of retaining only the largest nodes
in the network can be an obstacle to scaling up to very large corpora of
several million documents. Indeed, if we wish to concentrate on a few
hundred nodes, the threshold will be very high and the resulting network
risks being just a constellation of single nodes with no links between
them, the other nodes with which they are linked being in fact made
insignificant by the threshold set in terms of the number of
publications (or citations) per node. In addition, the processing time
for a very large corpus of publications can be very long, making such a
tool unusable in a web application where the user expects rapid
interaction with the application.

\hypertarget{network-analysis-at-scale}{%
\section{2. Network analysis at scale}\label{network-analysis-at-scale}}

We propose a method for overcoming the limitations set out above. We
also use a filtering technique to reduce the size of the network, but
with a dual approach: instead of filtering the nodes, we filter the
links.

\hypertarget{focusing-on-strongest-interactions}{%
\subsection{2.1 Focusing on strongest
interactions}\label{focusing-on-strongest-interactions}}

One of the added values of mapping with a network view is to show the
interactions between entities, i.e.~the links between the nodes in the
graph. These links provide crucial information that can be used to
structure communities. Here if the size of the network needs to be
reduced (for reasons of computation, speed, legibility and
interpretability), it is vital to preserve the links that carry the most
information, i.e.~the strongest interactions. With this reasoning, it
seems logical to reduce the size of the network by only affecting the
strongest links.

Here, we assume that there are no large isolated nodes (with no
connection). An entity with no connection will not appear in the
mapping. This assumption can sometimes prove to be false, particularly
in the case of authors in literature, for example.

Thus, from a given corpus, however large, we seek to extract the pairs
of entities with the strongest interactions, for example the most
co-signatures per pair of authors. From this list of pairs, we can
naturally find the nodes of the graph and deduce a new graph. If the
graph has several independent components, i.e.~several unconnected
sub-graphs, we can decide to keep only the main component(s).

\hypertarget{publication-metadata-enrichment-to-produce-different-mapping}{%
\subsection{2.2 Publication metadata enrichment to produce different
mapping}\label{publication-metadata-enrichment-to-produce-different-mapping}}

Each publication in the scanR corpus goes through a systematic
enrichment pipeline, including author and affiliation disambiguation,
full-text parsing and topic detection.

For authors, the French-specific persistent identifier (PID)
\url{https://www.idref.fr} is used. Its coverage, even if not perfect,
for French affiliated authors is strong thanks to the deep linking
between idref and the PhD thesis registration in France. Specific
heuristics have been implemented to disambiguate names and link them to
idref.

For affiliations, again French specific PID are used, especially
\url{https://sirene.fr} and \url{http://rnsr.fr}. SIRENE is a national
(French) PID for public and private institutions. RNSR is a French PID
for the research structures like laboratories. A specific module based
on Elasticsearch \url{https://github.com/dataesr/affiliation-matcher}
has been implemented to automatically link publications to those PIDs
(L'Hôte and Jeangirard 2021).

For topics, wikidata identifiers have been linked using the
entity-fishing module \url{https://github.com/kermitt2/entity-fishing}
(Foppiano and Romary 2020).

Other enrichments, like software detection are also present. These are
based on software mentions detections using GROBID and Softcite at scale
on the French publications corpus (Bassinet et al. 2023).

\hypertarget{elasticsearch-implementation}{%
\subsection{2.3 Elasticsearch
implementation}\label{elasticsearch-implementation}}

To identify the strongest links, it would be too costly to go through
the entire corpus. We have pre-calculated the links at the level of each
publication. So, if a publication is linked to 3 themes, T1, T2 and T3,
a pre-calculated field, at publication level, contains all T1-T2, T1-T3
and T2-T3 pairs. This co\_topics field represents the co-appearance
links within the publication. We then use elasticsearch's aggregation
functionality to list the most present links, very efficiently. By
default, we limit ourselves to the top 2000 links to ensure optimal
performance.

At the publication level, the pairs are calculated for authors,
institutions, laboratories, software, fundings and countries too.

In practice, a PID is also stored (the wikidata for topics, for example)
to disambiguate entities. In practice, for a given query, elasticsearch
returns a response containing the strongest links, for example:

\begin{Shaded}
\begin{Highlighting}[]
                \FunctionTok{\{}
                    \DataTypeTok{"key"}\FunctionTok{:} \StringTok{"Q15305550\#\#\#carbon sequestration{-}{-}{-}Q7942\#\#\#climate change"}\FunctionTok{,}
                    \DataTypeTok{"doc\_count"}\FunctionTok{:} \DecValTok{17}\FunctionTok{,}
                \FunctionTok{\}}\ErrorTok{,}
                \FunctionTok{\{}
                    \DataTypeTok{"key"}\FunctionTok{:} \StringTok{"Q15305550\#\#\#carbon sequestration{-}{-}{-}Q623\#\#\#carbon"}\FunctionTok{,}
                    \DataTypeTok{"doc\_count"}\FunctionTok{:} \DecValTok{14}\FunctionTok{,}
                \FunctionTok{\}}\ErrorTok{,}
                \FunctionTok{\{}
                    \DataTypeTok{"key"}\FunctionTok{:} \StringTok{"Q15305550\#\#\#Carbon sequestration{-}{-}{-}Q7942\#\#\#Climate change"}\FunctionTok{,}
                    \DataTypeTok{"doc\_count"}\FunctionTok{:} \DecValTok{13}\FunctionTok{,}
                \FunctionTok{\}}\ErrorTok{,}
                \FunctionTok{\{}
                    \DataTypeTok{"key"}\FunctionTok{:} \StringTok{"Q15305550\#\#\#Carbon sequestration{-}{-}{-}Q898653\#\#\#Climate change mitigation"}\FunctionTok{,}
                    \DataTypeTok{"doc\_count"}\FunctionTok{:} \DecValTok{10}\FunctionTok{,}
                \FunctionTok{\}}\ErrorTok{,}
                \FunctionTok{\{}
                    \DataTypeTok{"key"}\FunctionTok{:} \StringTok{"Q397350\#\#\#agroforestry{-}{-}{-}Q8486\#\#\#coffee"}\FunctionTok{,}
                    \DataTypeTok{"doc\_count"}\FunctionTok{:} \DecValTok{10}\FunctionTok{,}
                \FunctionTok{\}}\ErrorTok{,}
                \FunctionTok{\{}
                    \DataTypeTok{"key"}\FunctionTok{:} \StringTok{"Q15305550\#\#\#Carbon sequestration{-}{-}{-}Q1997\#\#\#CO2"}\FunctionTok{,}
                    \DataTypeTok{"doc\_count"}\FunctionTok{:} \DecValTok{9}\FunctionTok{,}
                \FunctionTok{\}}\ErrorTok{,}
                \FunctionTok{\{}
                    \DataTypeTok{"key"}\FunctionTok{:} \StringTok{"Q623\#\#\#carbon{-}{-}{-}Q627\#\#\#nitrogen"}\FunctionTok{,}
                    \DataTypeTok{"doc\_count"}\FunctionTok{:} \DecValTok{9}\FunctionTok{,}
                \FunctionTok{\}}\ErrorTok{,}
                \FunctionTok{\{}
                    \DataTypeTok{"key"}\FunctionTok{:} \StringTok{"Q15305550\#\#\#Carbon sequestration{-}{-}{-}Q623\#\#\#carbon"}\FunctionTok{,}
                    \DataTypeTok{"doc\_count"}\FunctionTok{:} \DecValTok{7}\FunctionTok{,}
                \FunctionTok{\}}\ErrorTok{,}
\end{Highlighting}
\end{Shaded}

\hypertarget{network-creation}{%
\subsection{2.4 Network creation}\label{network-creation}}

The network creation process involves several key steps: transforming
Elasticsearch results into a graph, filtering the network to focus on
the most interesting nodes, applying spatialization algorithms for
visualization, and detecting communities within the network. Below, we
detail each of these steps.

The network creation process begins with the results obtained from
Elasticsearch, utilizing the open-source JavaScript library Graphology
\url{https://github.com/graphology/graphology} to construct and
manipulate the network. Each link result from Elasticsearch is
transformed into nodes and edges, with edge strength corresponding to
the number of aggregated documents.

To ensure that the network remains manageable and focuses on the most
interesting nodes, we employ a strategy that prioritizes the
best-connected nodes rather than the largest nodes. By default, the
maximum number of nodes is set to 300. This threshold helps in
maintaining the computational efficiency and interpretability of the
network.

In graph theory, a component refers to a subgraph in which any two nodes
are connected to each other by paths, and which is connected to no
additional nodes in the larger graph. Using Graphology, we filter the
network components by iteratively removing the smallest components until
the number of nodes falls below the threshold or only one component
remains. This largest component is then subjected to further filtering
if it still exceeds the node threshold. In this second filtering step,
we utilize the betweenness centrality metric to retain the
best-connected nodes. Betweenness centrality measures the extent to
which a node lies on the shortest path between other nodes, thereby
identifying nodes that act as bridges within the network.

Once the filtering process is complete, we apply a spatialization
algorithm to position the nodes in a 2D space. For this purpose, we use
the ForceAtlas2 algorithm, which is designed to produce informative and
aesthetically pleasing layouts by simulating a physical system where
nodes repel each other and edges act as springs pulling connected nodes
together. This results in a clear and intuitive visual representation of
the network (Jacomy 2014). Thanks to Graphology, the settings of the
ForceAtlas2 algorithm are automatically infered from our network order
(number of nodes) as below:

\begin{verbatim}
barnesHutOptimize: order > 2000,
strongGravityMode: true,
gravity: 0.05,
scalingRatio: 10,
slowDown: 1 + Math.log(order)
\end{verbatim}

In graph theory, a community corresponds to a set of nodes in a graph
that are strongly interconnected with each other, while being less
connected with nodes outside this community. Communities can be
identified in order to understand the underlying structure and patterns
of the graph, as well as to analyze the relationships and interactions
between the entities that make it up. To identify and visualize
communities within the network, we apply the Louvain algorithm using
Graphology. This algorithm works by optimizing a modularity measure that
evaluates the strength of communities in a graph (Blondel et al. 2008).
More precisely, Louvain seeks to maximize modularity by progressively
moving the nodes of a graph into different communities, in an iterative
fashion. At each stage, he merges neighboring communities if this leads
to an improvement in the overall modularity of the graph. This iterative
process continues until no further moves can increase modularity.

The \texttt{graphology-communities-louvain} node module is being used.
This way, each step (like spatizalization, community-detection) are
implemented modularly. A benchmark, in our use case, of the Louvain and
the Leiden algorithms would be desirable. The graphology library started
a while ago working on an implementation of the Leiden algorithm (see
\url{https://github.com/graphology/graphology/tree/master/src/communities-leiden})
but that remains to be implemented.

\hypertarget{vosviewer-implementation}{%
\subsection{2.5 VOSviewer
implementation}\label{vosviewer-implementation}}

To display the network within our application, we use the open source
VOSviewer online tool for network visualization
\url{https://github.com/neesjanvaneck/VOSviewer-Online}. It is based on
the VOSviewer software which is very popular for network analysis in
bibliometric studies (Waltman, Eck, and Noyons 2010).

VOSviewer accepts JSON files formatted according to a specific template
\url{https://app.vosviewer.com/docs/file-types/json-file-type}. This
template includes essential attributes for nodes and edges, such as the
node ID, label, position (x, y), and additional metadata. To ensure
compatibility, we transform our Graphology object into a JSON file that
complies to VOSviewer's required format.

Once the JSON file is generated, VOSviewer renders the network,
displaying nodes and edges in an interactive and visually appealing
manner. The nodes are colorized based on the communities identified
through the clustering process performed using the Louvain algorithm.
This colorization helps in visually distinguishing different communities
within the network, making it easier to analyze and interpret the
underlying structure and interactions.

VOSviewer includes its own spatialization algorithm and parameters for
layout customization. However, after testing these options, we found
them to be visually less intuitive and informative. Consequently, we
chose to use the ForceAtlas2 algorithm for spatialization, as described
in the previous section, which offers a more aesthetically pleasing and
informative layout by being automatically set for our network.

\begin{figure}
\centering
\includegraphics{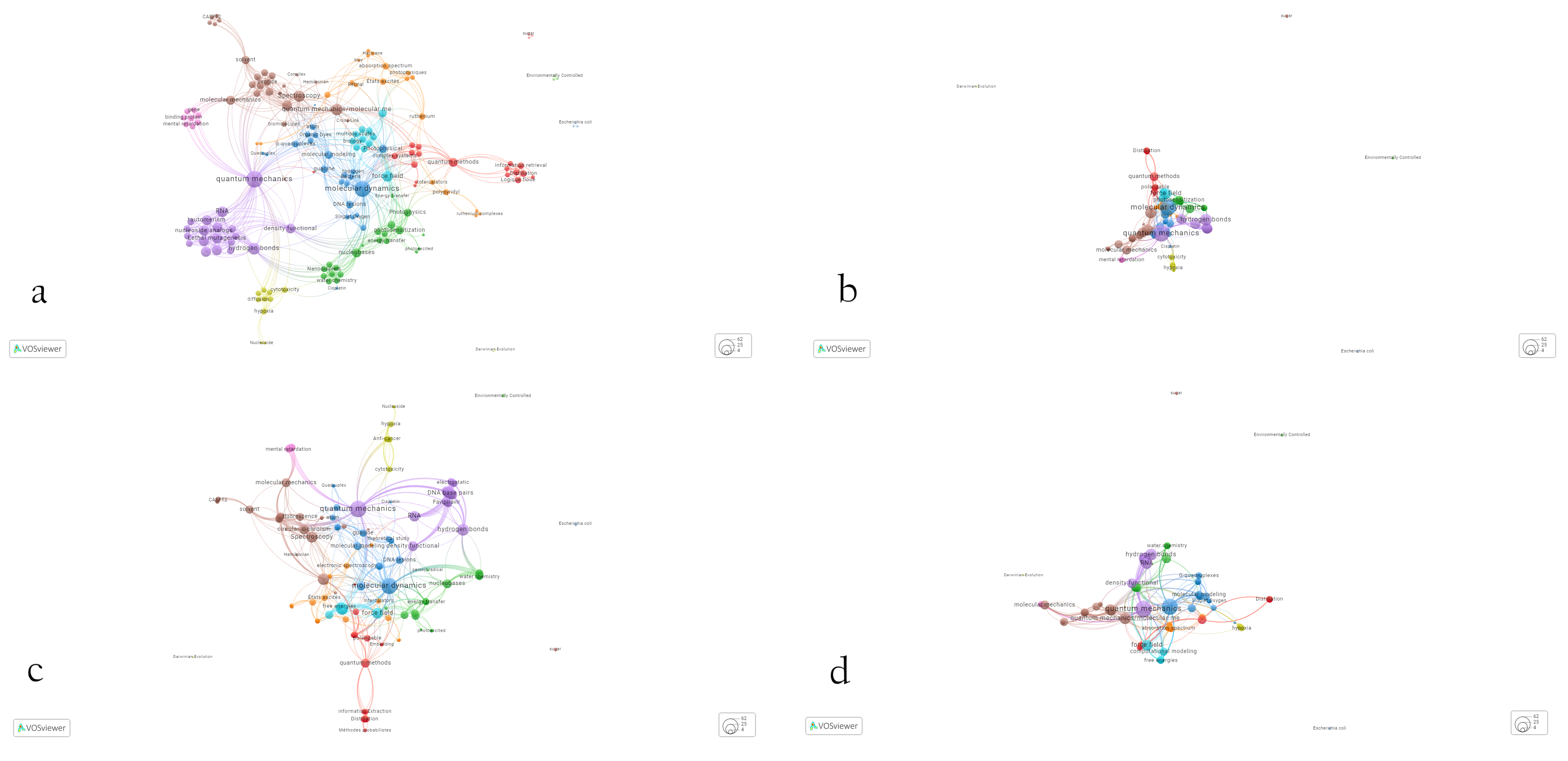}
\caption{Visualization of a network with VOSviewer.\\
\emph{(a) Using ForceAltlas2 spatialization with infered settings (b)
Using VOSviewer spatialization (attraction=2, repulsion=1) (c) Using
VOSviewer spatialization (attraction=3, repulsion=1) (d) Using VOSviewer
spatialization (attraction=1, repulsion=0)}}
\end{figure}

\hypertarget{making-insightful-maps}{%
\section{3. Making insightful maps}\label{making-insightful-maps}}

This scanR feature is designed to help users gain a better understanding
of the underlying structures via thematic or co-publication maps. To
help the user, it's important to be able to characterize each of the
communities automatically identified. It is therefore important to label
each community before describing them.

\hypertarget{llm-trick}{%
\subsection{3.1 LLM trick}\label{llm-trick}}

To name the communities we use generative AI from Mistral AI
(`open-mistral-nemo' model). The names are obtained from the main themes
of the publications collected for each community. For the time being, we
limit ourselves to the 2000 most relevant publications (in relation to
the user's search) for each community. The following prompt is used:

\begin{quote}
`` You have been tasked with naming distinct fields of study for several
communities of research publications. Below are lists of topics and
their weights representing each community. Your goal is to provide a
unique and descriptive name for each field of study that best
encapsulates the essence of the topics within that community. Each
should be unique and as short as possible. If the list of topic is
empty, output a empty string. Output as JSON object with the list number
and the single unique generated name. ''
\end{quote}

To illustrate its functionality, consider the following example:

\begin{verbatim}
// Input with each list corresponding to a community
"list1 = [Soil (8), Carbon Sequestration (5), Soil Organic Matter (5), Carbon (5),  
Ecosystem Services (5), Priming Effect (4), Sequestration (4), Amazonian (3), Andosol (3)],  
list2 = [Soil Organic Carbon (11), Carbon (10), Climate Change (7), Soil (7),  
Carbon Sequestration (6), Carbon Cycle (5), Soil Carbon (4)],  
list3 = [Acl (7), Carbon (3), Carbon Sequestration (3), South Pacific Ocean (3),  
Trichodesmium (3), Crocosphaera (2), Crocosphaera-watsonii (2), Dinitrogen-fixation (2)]"

// Mistral output
\end{verbatim}

\begin{Shaded}
\begin{Highlighting}[]
\FunctionTok{\{}
  \DataTypeTok{"list1"}\FunctionTok{:} \StringTok{"Amazon Andosol Carbon Dynamics"}\FunctionTok{,}
  \DataTypeTok{"list2"}\FunctionTok{:} \StringTok{"Soil Carbon and Climate Change"}\FunctionTok{,}
  \DataTypeTok{"list3"}\FunctionTok{:} \StringTok{"South Pacific Ocean Carbon Cycling"}
\FunctionTok{\}}
\end{Highlighting}
\end{Shaded}

\hypertarget{citation-hot-topics}{%
\subsection{3.1 Citation / hot topics}\label{citation-hot-topics}}

A citation score is estimated for each cluster. This score relates the
number of recent citations (over the last two years) to the number of
total publications in the cluster. This score is intended to help detect
hotspots in the communities identified in the corpus. We use citations
data from OpenAlex \url{https://openalex.org/}, which is as of today one
of the best open source bibliometric datasource. However, citations
metadata from OpenAlex remains incomplete and must therefore be
interpreted with caution (Alperin et al. 2024).

\hypertarget{custom-perimeter}{%
\subsection{3.2 Custom perimeter}\label{custom-perimeter}}

scanR offers this mapping tool for the entire indexed corpus, but it is
also possible to adapt the tool to a restricted perimeter, at the user's
discretion. For example, an institution or laboratory can define its own
corpus (based on a list of publications) and a mapping tool dedicated to
this perimeter is automatically created. Technically, elasticsearch
queries are the same, with just an additional filter to query only the
publications within the perimeter. The tool can be embedded in any
website using an iframe. It's the same principle as the local Open
Science Monitor: any French institution can benefit from the whole
infrastructure already inplace and get a custom tool based on the same
data, treatments and technologic stack as the national tool. This
approach eliminates the need for automatic alignment of affiliations,
which remains a highly complex task. Automation is possible to a certain
extent (L'Hôte and Jeangirard 2021), but human curation remains
necessary in the majority of cases (Jeangirard, Bracco, and L'Hôte
2024). In this way, users retain control over the definition of their
perimeter, and can, if they wish, have several distinct perimeters.

\hypertarget{code-availibility}{%
\section{4. Code availibility}\label{code-availibility}}

The code developed for the scanR web application is open source and
available online on GitHub \url{https://github.com/dataesr/scanr-ui}
under MIT license.

\hypertarget{references}{%
\section*{References}\label{references}}
\addcontentsline{toc}{section}{References}

\hypertarget{refs}{}
\begin{cslreferences}
\leavevmode\hypertarget{ref-alperin2024analysissuitabilityopenalexbibliometric}{}%
Alperin, Juan Pablo, Jason Portenoy, Kyle Demes, Vincent Larivière, and
Stefanie Haustein. 2024. ``An Analysis of the Suitability of Openalex
for Bibliometric Analyses.'' \url{https://arxiv.org/abs/2404.17663}.

\leavevmode\hypertarget{ref-bassinet:hal-04121339}{}%
Bassinet, Aricia, Laetitia Bracco, Anne L'Hôte, Eric Jeangirard, Patrice
Lopez, and Laurent Romary. 2023. ``Large-scale Machine-Learning analysis
of scientific PDF for monitoring the production and the openness of
research data and software in France.''
\url{https://hal.science/hal-04121339}.

\leavevmode\hypertarget{ref-Blondel_2008}{}%
Blondel, Vincent D, Jean-Loup Guillaume, Renaud Lambiotte, and Etienne
Lefebvre. 2008. ``Fast Unfolding of Communities in Large Networks.''
\emph{Journal of Statistical Mechanics: Theory and Experiment} 2008
(10): P10008. \url{https://doi.org/10.1088/1742-5468/2008/10/P10008}.

\leavevmode\hypertarget{ref-10.1162ux2fqss_a_00179}{}%
Chaignon, Lauranne, and Daniel Egret. 2022. ``Identifying Scientific
Publications Countrywide and Measuring Their Open Access: The Case of
the French Open Science Barometer (Bso).'' \emph{Quantitative Science
Studies} 3 (1): 18--36. \url{https://doi.org/10.1162/qss_a_00179}.

\leavevmode\hypertarget{ref-foppiano2020entity}{}%
Foppiano, Luca, and Laurent Romary. 2020. ``Entity-Fishing: A Dariah
Entity Recognition and Disambiguation Service.'' \emph{Journal of the
Japanese Association for Digital Humanities} 5 (1): 22--60.

\leavevmode\hypertarget{ref-10.1371ux2fjournal.pone.0098679}{}%
Jacomy, Tommaso AND Heymann, Mathieu AND Venturini. 2014. ``ForceAtlas2,
a Continuous Graph Layout Algorithm for Handy Network Visualization
Designed for the Gephi Software.'' \emph{PLOS ONE} 9 (6): 1--12.
\url{https://doi.org/10.1371/journal.pone.0098679}.

\leavevmode\hypertarget{ref-jeangirard:hal-04813230}{}%
Jeangirard, Eric. 2024. ``scanR - Explore public data on French research
and innovation.'' In \emph{euroCRIS SMM 2024}. Paris, France: euroCRIS.
\url{https://hal.science/hal-04813230}.

\leavevmode\hypertarget{ref-jeangirard:hal-04598201}{}%
Jeangirard, Eric, Laetitia Bracco, and Anne L'Hôte. 2024. ``Works-magnet
: aucune de perdue, 10 000 de retrouvées.'' Abes; Journées Abes 2024.
\url{https://doi.org/10.5281/zenodo.11471247}.

\leavevmode\hypertarget{ref-lhote_using_2021}{}%
L'Hôte, Anne, and Eric Jeangirard. 2021. ``Using Elasticsearch for
Entity Recognition in Affiliation Disambiguation.''
\emph{arXiv:2110.01958 {[}Cs{]}}, October.
\url{http://arxiv.org/abs/2110.01958}.

\leavevmode\hypertarget{ref-DBLP:journalsux2fcorrux2fabs-1006-1032}{}%
Waltman, Ludo, Nees Jan van Eck, and Ed C. M. Noyons. 2010. ``A Unified
Approach to Mapping and Clustering of Bibliometric Networks.''
\emph{CoRR} abs/1006.1032. \url{http://arxiv.org/abs/1006.1032}.

\leavevmode\hypertarget{ref-DBLP:booksux2fcuux2fWB2021}{}%
Wang, Dashun, and Albert-László Barabási. 2021. \emph{The Science of
Science}. Cambridge University Press.
\url{https://doi.org/10.1017/9781108610834}.
\end{cslreferences}

\end{document}